\pdfoutput=1
\documentclass[aps,twocolumn,a4paper,showpacs,superscriptaddress,
    floatfix]{revtex4-1}
\usepackage{amsmath,amssymb}
\usepackage[hypertexnames=false]{hyperref}
\usepackage{datetime}
\usepackage{enumitem}
\usepackage{graphicx}
\usepackage{siunitx}
\usepackage{braket}
\usepackage{esdiff}

\usepackage[dvipsnames]{xcolor}
\usepackage{xfrac} %sfrac
\usepackage{tikz}

\DeclareSIUnit\gauss{G}

\hypersetup{
    colorlinks,
    linkcolor={red!50!black},
    citecolor={blue!50!black},
    urlcolor={blue!80!black}
}

\begin{document}

\title{Quantized conductance through a dissipative atomic point contact}
\author{Laura Corman}
\author{Philipp Fabritius}
\author{Samuel H\"{a}usler}
\author{Jeffrey Mohan}
\affiliation{Department of Physics, ETH Zurich, 8093 Z\"{u}rich, Switzerland}
\author{Lena H. Dogra}
\affiliation{Department of Physics, ETH Zurich, 8093 Z\"{u}rich, Switzerland}
\affiliation{Cavendish Laboratory, University of Cambridge, J. J. Thomson Avenue, Cambridge CB3 0HE, United Kingdom}
\author{Dominik Husmann}
\author{Martin Lebrat}
\author{Tilman Esslinger}
\affiliation{Department of Physics, ETH Zurich, 8093 Z\"{u}rich, Switzerland}

% \date{\today, \currenttime}

\begin{abstract}
Signatures of quantum transport are expected to quickly vanish as dissipation is introduced in a system. This dissipation can take several forms, including that of particle loss, which has the consequence that the total probability current is not conserved.
Here, we study the effect of such losses at a quantum point contact (QPC) for ultracold atoms. 
Experimentally, dissipation is provided by a near-resonant optical tweezer whose power and detuning control the loss rates for the different internal atomic states as well as their effective Zeeman shifts. 
We theoretically model this situation by including losses in the Landauer-B\"uttiker formalism over a wide range of dissipative rates. 
We find good agreement between our measurements and our model, both featuring robust conductance plateaus. 
Finally, we are able to map out the atomic density by varying the position of the near-resonant tweezer inside the QPC, realizing a dissipative scanning gate microscope for cold atoms.

\end{abstract}

\maketitle

\section{Introduction}

Coupling a system to its environment is a central concept in physics, giving rise to different types of ensembles in thermodynamics \cite{balian2007microphysics}. It leads to fundamental questions about the fate of quantum mechanics at a macroscopic scale \cite{schrodinger1935present, raimond2006exploring}. The coupling to the environment can also compete with coherence and interaction effects, leading to new phenomena \cite{muller2012engineered, daley2014quantum}. 
A full model of a system and its environment is out of reach in most cases due to the exponential growth of the total Hilbert space with the number of degrees of freedom. 
Tracing out those of the environment can be done under certain assumptions leading to a Lindblad master equation \cite{breuer2002theory}.
In some cases, further simplification is possible by only adding dissipative terms to the isolated system's equation of motion \cite{caldeira1983quantum,leggett1987dynamics}. 
The resulting dynamics, which becomes non-Hermitian, has recently found renewed interest due to the exotic behaviors of exceptional points linked to the collapse of the eigenvectors at a critical dissipation strength \cite{moiseyev_2011, miri2019exceptional}.

Dissipation, understood as the non-conservation of the system's volume in phase space over time, is characteristic for transport phenomena where currents bring a system towards equilibrium while producing entropy \cite{balian2007microphysics}. 
There, extrinsic dissipation can be included by physical or fictitious coupling to a reservoir other than the leads driving the transport processes.  
Such an additional coupling has mostly been studied to model incoherent scattering of electrons \cite{buttiker1986role, datta1989steady, sols1992scattering}. 
It is usually done by adding a fictitious reservoir with which the lossy region can exchange particles without any net current \cite{buttiker1986role}. 
Mimicking particle losses as a perturbation to transport by an additional absorbing reservoir has to our knowledge so far not been explored in this context. 

Cold clouds of atoms are nearly closed systems by default, but can be opened by designing atom losses in a controlled way.
Experimental loss channels include (i) molecule formation via photoassociation \cite{tomita2017observation} or via decay to a molecular channel \cite{syassen2008strong, amico2018time, schemmer2018cooling}, (ii) ionization using a focused electron beam \cite{gericke_high-resolution_2008, barontini_controlling_2013} or a femtosecond laser \cite{wessels2018absolute} or (iii) scattering of near-resonant photons that impart a large kinetic energy to the atoms \cite{pfau1994loss, patil2015measurement, bouganne2019anomalous}. 
Many of these techniques realize localized losses, an assumption underlying the usual theoretical models for dissipation in transport structures \cite{sols1992scattering}, realized using an electron beam in a Bose-Einstein condensate \cite{labouvie2016bistability, mullers2018coherent}. 

\begin{figure}
    \includegraphics[scale=1.0]{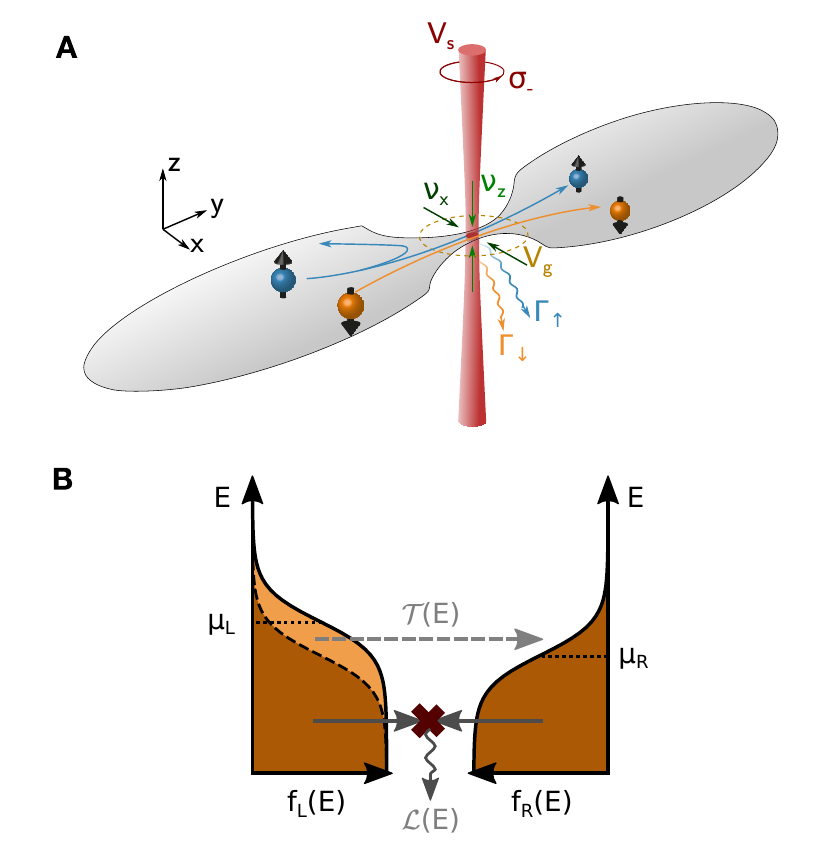}
    \caption{\textbf{Experimental realization of a Landauer-B\"uttiker setup with losses.} 
    (\textbf{a}) $^6$Li atoms in the lowest ($\ket{\downarrow}$, orange) and third-lowest  ($\ket{\uparrow}$, blue) hyperfine states are allowed to flow between two reservoirs connected by an optically-defined QPC (grey).  A near-resonant optical tweezer (red) with waist $w_s = \SI{2.0(1)}{\micro\meter}$ introduces different loss rates $\Gamma_{\uparrow}$ and $\Gamma_{\downarrow}$ as well as a spin-dependent potential $V_s$ inside the QPC.
    A far-detuned gate beam (dashed circle) locally increases the chemical potential $\mu_\text{res}$ imposed by the reservoirs by $V_g$.    
    (\textbf{b}) In a Landauer-B\"{u}ttiker picture and in the absence of a temperature bias, net transport can be attributed to the excess of particles in the left reservoir not compensated by the right reservoir (orange)
    due to a chemical potential difference $\mu_L - \mu_R > 0$.
    Losses on the other hand involve the energy integration of the reservoir Fermi-Dirac distributions $f_L(E)$ and $f_R(E)$, 
    including states below the Fermi level that do not contribute to a net current (brown). 
    Conductance and loss are furthermore weighted by the energy-dependent transmission $\mathcal{T}(E)$ and loss $\mathcal{L}(E)$ coefficients of the mesoscopic channel.
    }
    \label{fig:setup}
\end{figure}

In this paper, we study dissipation as a perturbation to transport through a quantum point contact (QPC) for a two-component fermionic gas of ultracold $^6$Li atoms, where the different hyperfine states of the atoms are interpreted as a pseudo-spin. 
The geometry of the optical potentials trapping the atoms realizes a two-terminal transport setup as illustrated in Fig~\ref{fig:setup}a  \cite{krinner_two-terminal_2017}. There, a cloud of degenerate fermionic lithium is divided into two reservoirs connected by a quasi one-dimensional (1D) constriction formed by two intersecting beams with a nodal line. Controlled by the chemical potential inside the 1D channel, only one to two transverse modes are available to particles moving from one reservoir to the other. Since the potential landscape in the channel is smooth, transport is ballistic: the transmission probability of the atoms is close to unity. This leads to the measurement of conductance plateaus \cite{krinner_observation_2015} as the chemical potential is increased in the wire, in agreement with the Landauer-B\"uttiker model (illustrated in Fig~\ref{fig:setup}b) that identifies conductance with transmission for mesoscopic conductors \cite{landauer1957spatial, buttiker1986four, imry2002introduction}.

The losses are provided by a near-resonant beam focused onto the constriction. The spatial profile of the beam is controlled holographically using a digital micromirror device, allowing us to correct for aberrations \cite{zupancic2016ultra}. 
The scattering rate and dipole potential experienced by the atoms are related to the imaginary and real part of the atomic polarizability, respectively, which must be computed in the regimes of high magnetic fields relevant for tuning the scattering properties of $^6$Li.
For light frequencies close to the atomic resonances of different pseudo-spin states, both scattering rate and dipole potential are strongly spin-dependent due to the splitting of their transition frequencies.
The dipole potential can be interpreted as an effective Zeeman shift in the case where its magnitudes are equal and opposite for both spins.
This case is investigated in our companion paper \cite{Lebrat2019companion} wherein the amplitude of this effective Zeeman shift is on the order of the Fermi energy of the atoms, leading to a significant shift between the onset of the conductance plateaus of the two spins.
Here, we focus on the effect of the atom losses engineered for the atoms. 
We demonstrate that in spite of the dissipation, the conductance plateaus persist owing to the constant flow of low temperature particles through the channel.
We show that the transport properties of a lossy QPC can be described by including the transmission as well as the energy-dependent losses characterizing the channel into a Landauer-B\"uttiker model as illustrated in Fig~\ref{fig:setup}b.
Finally, we are able to reconstruct the atomic density in a wide region around the QPC by monitoring the atom losses as the position of the near-resonant tweezer is varied. 

This paper is structured as follows. In Sec.~\ref{sec:polarizability}, we review the polarizability of alkali atoms in high magnetic fields. In Sec.~\ref{sec:lossy_landauer}, we adapt the Landauer formalism to the presence of particle losses. In Sec.~\ref{sec:loss_mechanisms}, we describe the experimental setup and the loss mechanims induced by the near-resonant light, justifying the validity of the analysis of Sec.~\ref{sec:lossy_landauer}. Finally, we compare the extended Landauer formalism to our measurements in Sec.~\ref{sec:experiment}.

\section{$^6$Li polarizability in high magnetic fields \label{sec:polarizability}}

In the following, we briefly review how to compute the atomic polarizability in presence of a high magnetic field, which is particularly relevant when the interaction strength can be controlled using a Feshbach resonance such as $^6$Li. Choosing the frequency and intensity of a near-resonant beam to tailor different optical potentials and losses for each internal state of ultracold lithium atoms. We will focus in presenting regimes where the optical potential is of equal magnitude but opposite sign for the two relevant states in the experiment, emulating an effective Zeeman shift. 

\subsection{Polarizability of alkali atoms in the Paschen-Back regime}

We consider alkali atoms in a uniform magnetic field $\vec{B}$. If the energy of the electron's magnetic dipole moment exceeds the hyperfine interaction for $^6$Li, the atoms are in the Paschen-Back regime.

The Hamiltonian for a single atom can be expressed as a function of the electronic and nuclear spin operators $\hat{\vec{J}} = \hat{\vec{L}} + \hat{\vec{S}}$ and $\hat{\vec{I}}$
\begin{equation}
\hat{H}_\text{at} = \frac{\mu_B}{\hbar} (g_J \hat{\vec{J}} \cdot \vec{B} + g_I \hat{\vec{I}} \cdot \vec{B}) + \frac{a}{\hbar^2} \hat{\vec{J}} \cdot \hat{\vec{I}}, \label{eq:hamiltonian}
\end{equation}
where $\mu_B$ is the Bohr magneton, $g_J$ and $g_I$ the electronic and nuclear gyromagnetic factors, and $a$ the hyperfine coupling energy. Both $g_J$ and $a$ depend on the electron's orbital momentum $L$ and total spin $J$ which characterize each level manifold $2 S$, $2 P_{1/2}$, and $2 P_{3/2}$. Their values for $^6$Li are reported in \cite{gehm_2003}.

The dipole matrix element between two states labeled by electron and nuclear spin $(J, m_J, m_I)$ is given by the Wigner-Eckardt theorem:
\begin{equation}
d^q_{k k'} = \frac{1}{\sqrt{2 J' + 1}} \delta_{m_I, m_I'} \braket{J m_J, 1 q|J' m_J'} \braket{J'||\vec{d}||J}, \label{eq:dipole_matrix_element}
\end{equation}
where $\braket{J m_J, 1 q|J' m_J'}$ is the Clebsch-Gordan coefficient relating initial and final electron spins $m_J$ and $m_J'$ through the added angular momentum $q = -1, 0, +1$ of a photon with
polarization $\sigma^-, \pi, \sigma^+$, respectively, and $\braket{J'||\vec{d}||J}$ is the reduced matrix element of the corresponding ${2S} \rightarrow {2P}$ transition.
The dipole matrix elements $d^{\, q}_{eg}$ between each pair of eigenstates $\ket{g}$ and $\ket{e}$ of (\ref{eq:hamiltonian}) are then given via a change of basis.

These dipole matrix elements appear in the complex polarizability tensor of each state $\ket{g}$ of the ground manifold $2S$ at light frequency $\nu$:
\begin{equation}
\alpha^{q q'}_g = -\sum_e \frac{(d^{\, q}_{eg})^* d^{\, q'}_{eg}}{h(\nu-\nu_{eg}) + i \hbar \Gamma_{eg}/2}\label{eq:polarizability}, 
\end{equation}
where the sum runs over all excited states $\ket{e}$ in $2 P_{1/2}$ and $2 P_{3/2}$. The transition frequencies $\nu_{eg}$ between ground and excited states are derived from the eigenvalues of (\ref{eq:hamiltonian}), and $\Gamma_{eg}$ denotes the spontaneous emission rates from excited to ground states:
\begin{equation}
\Gamma_{eg} = \frac{(h \nu_{eg})^3}{3 \pi \epsilon_0 c^3 \hbar^4} \sum_q |d^{\, q}_{eg}|^2,
\end{equation}
where $c$ is the speed of light and $\epsilon_0$ is the vacuum permittivity.

For light intensities $I$ small compared to the saturation intensity, the light shift $V_{\ket g}$ and scattering rate $\Gamma_{\ket g} $ experienced in the ground state $\ket{g}$ are related to the dispersive and dissipative parts of the polarizability:

\begin{align}
    V_{\ket g} = -\frac{I}{2 \epsilon_0 c} &\sum_{q, q'} e_q \, \text{Re}(\alpha^{q q'}_g) \, e_{q'}^* \label{eq:dipole_potential} \\  
    \Gamma_{\ket g} = \frac{I}{\hbar \epsilon_0 c} &\sum_{q, q'} e_q  \, \text{Im}(\alpha^{q q'}_g) \, e_{q'}^* \label{eq:scattering_rate}.
\end{align}

The sum indicates a double tensor contraction by the light polarization $\vec{e} = \vec{E}/|\vec{E}| = \sum_q e_q \vec{u}_q$, whose coordinates are expressed here in the basis
$\vec{u}_{-1} = (\vec{u}_x - i\vec{u}_y)/\sqrt{2}$ for $\sigma^-$-polarization, $\vec{u}_0 = \vec{u}_z$ for $\pi$-polarization, and $\vec{u}_{+1} = (\vec{u}_x + i\vec{u}_y)/\sqrt{2}$ for $\sigma^+$-polarization.

\subsection{Spin filter regimes for ultracold $^6$Li atoms}

The eigenstates of the Hamiltonian (\ref{eq:hamiltonian}) can be expressed as linear combinations of the $|J, m_J, m_I\rangle$ states. From an experimental point of view, we are especially interested in the first and third  lowest hyperfine states as possible ground states, labelled as $ \ket{\downarrow}$ and $ \ket{\uparrow}$.

At high magnetic field $B$, these states can be approximated by 

\begin{align}
    \ket{\downarrow} = |m_J=-\frac 1 2, m_I=1\rangle - \frac{\epsilon}{\sqrt 2}|m_J=\frac 1 2, m_I=0\rangle  \label{state_1} \\   
    \ket{\uparrow} = |m_J=-\frac 1 2, m_I=-1\rangle  \label{state_3} 
\end{align}
to first order in $\epsilon=ah/(g_J\mu_B B)$, with $a = 152.1\,{\rm MHz}$ the hyperfine coupling constant of the $2^2{\rm S}$ manifold.

Therefore the different internal states are mostly described by their electronic spin $m_J= -1/2$ and nuclear spin, $m_I = +1$ and $m_I = -1$, for the lowest and third-lowest hyperfine states of $^6$Li, respectively.  

For a given polarization, the transitions between these hyperfine states and the excited manifolds $2P_{1/2}$ and $2P_{3/2}$ occur at different frequencies, typically offset by a magnetic-field dependent shift comparable to the hyperfine coupling.
Tuning the light frequency in the vicinity of these transitions leads to strongly state-dependent polarizabilities which can be exploited experimentally.

As an example, and along the lines of our companion paper \cite{Lebrat2019companion}, the differential light shift created by a near-resonant beam can be reinterpreted as an effective Zeeman shift. We therefore aim at creating the maximum potential difference $V_{\uparrow}-V_{\downarrow}$ for a limited scattering rate at a frequency where the mean potential for the two spin states vanishes $(V_{\downarrow}+V_{\uparrow})/2=0$. 

\begin{figure*}
    \centering
    \includegraphics[scale=1]{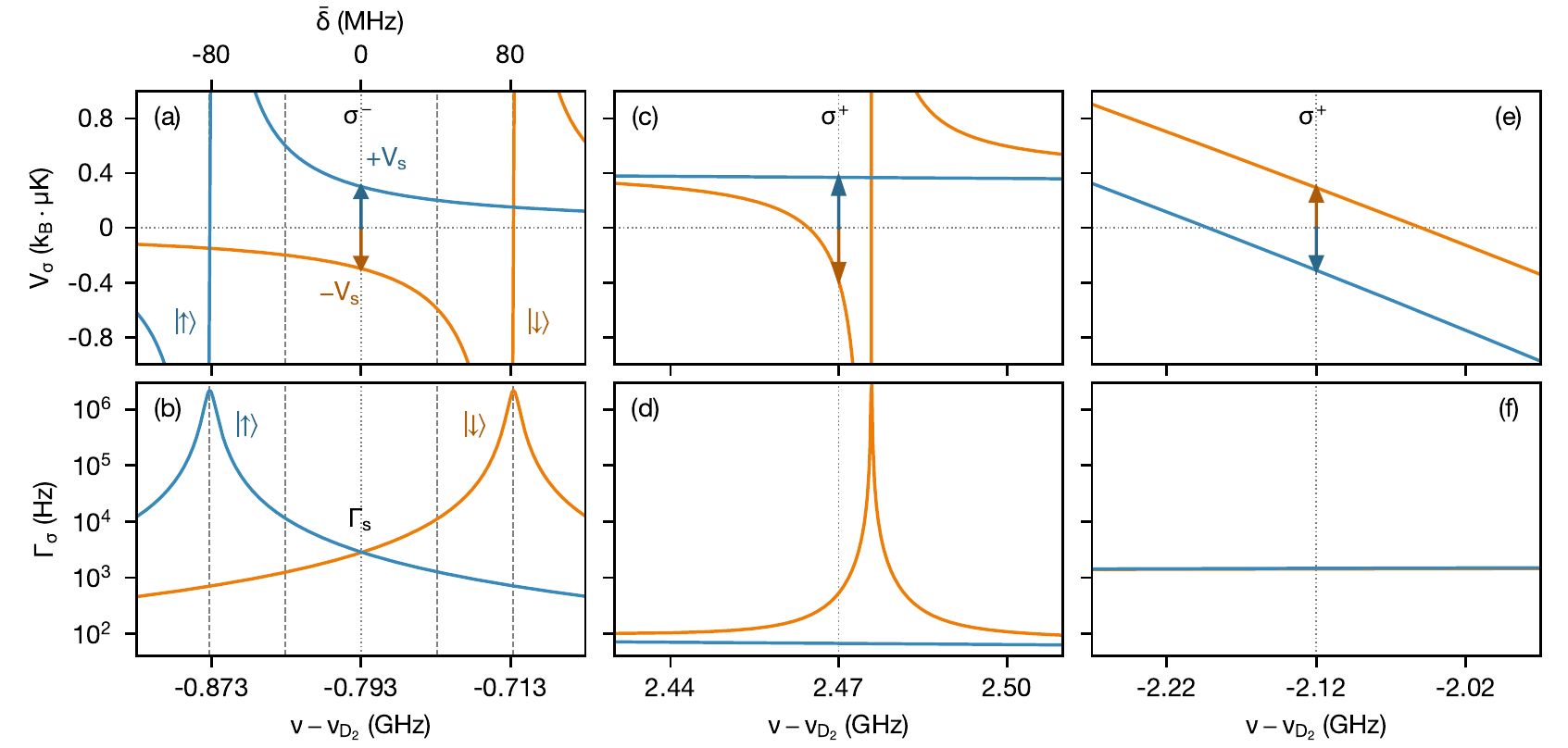}
    \caption{\textbf{Potentials and scattering rates at high magnetic field.}
    Potentials and scattering rates at a magnetic field of $B=568\,$G for the $\ket{\downarrow}$ and $\ket{\uparrow}$ states for different frequencies $\nu$ of the light where the mean potential $(V_{\downarrow}+V_{\uparrow})/2$ vanishes.  The lower frequency axes are offset by the $D_2$ transition frequency at zero magnetic field $\nu_{D_2}$ \cite{gehm_2003}. The intensities in each pair of graphs are chosen to give a constant amplitude of the spin potential $|V_{\downarrow}-V_{\uparrow}|=k_B\cdot \SI{0.6}{\micro\kelvin}$. 
    \textbf{a-b} correspond to $\sigma^-$ polarization and an intensity of $\SI{3}{\watt/\meter^2}$. The top frequency axis has been offset by the mean of the resonance frequencies of state $\ket{\downarrow}$ and $\ket{\uparrow}$ for the main $D_2$ $\sigma^-$ transition:  $\bar{\delta} = \nu -(\nu_{\downarrow}+\nu_{\uparrow})/2$  
    \textbf{c-d} correspond to $\sigma^+$ polarization and an intensity of $\SI{136}{\watt/\meter^2}$ close to a narrow $D_2$ resonance appearing in the Paschen-Pack regime, 
    \textbf{e-f} correspond to $\sigma^+$ polarization and an intensity of $1.2\cdot 10^4 \,\SI{}{\watt/\meter^2}$ between the $D_1$ and $D_2$ lines.
    }
    \label{fig:theory_detuning_high_B}
\end{figure*}

A natural and experimentally practical solution adopted in \cite{Lebrat2019companion} is to choose a frequency $\nu$ right in between the strongest transitions of the states $\ket{\downarrow}$ and $\ket{\uparrow}$ to the $D_2$ line, which fulfill the condition $m_{J'} = m_J + q$ and whose linewidths are almost equal at high magnetic field.
For $\sigma^-$ light, the resonance frequencies $\nu_{\downarrow}$ and $\nu_{\uparrow}$ are $ \SI{162.6}{\mega\hertz}$ apart at the typical magnetic field of $ B=\SI{568}{\gauss}$ where the scattering length between the two spin states vanishes.
The potentials and scattering rates for each atomic internal state are displayed on Fig.~\ref{fig:theory_detuning_high_B}ab as a function of the mean detuning $\bar{\delta} = \nu -(\nu_{\downarrow}+\nu_{\uparrow})/2$.

Alternative solutions are however more favorable in terms of light scattering.
Interestingly, the small admixture of $m_J=1/2$ states in $\ket{\downarrow}$ as seen in equation (\ref{state_1}). It leads to narrow transitions to the states $m_{J'}=1/2 + q$ with an effective width $\sim \epsilon^2\Gamma/2$.
Figure \ref{fig:theory_detuning_high_B}cd shows the dipole potential and scattering rate close to such a transition for $\sigma^+$-polarization within the $D_2$ line, adapting the light intensity such that the differential light shift remains equal to the one in Fig.~\ref{fig:theory_detuning_high_B}a.
The associated scattering rate is reduced by more than one order of magnitude compared to Fig.~\ref{fig:theory_detuning_high_B}b.

 Last, the mean potential also vanishes at a frequency between the $D_1$ and the $D_2$ lines for $\sigma^+$ and $\pi$ polarization. 
As illustrated in Fig.~\ref{fig:theory_detuning_high_B}ef for $\sigma^+$ polarization, the signs of the dipole potentials $V_\uparrow$ and $V_\downarrow$ are inverted compared to the two previous cases.
The scattering rate at fixed differential light shift is only reduced by a factor of two, but it is equal for the two spin states and remains weakly frequency-dependent.

\section{Adapting the Landauer formalism with losses\label{sec:lossy_landauer}}

Transport between two reservoirs through a QPC is usually well described by Landauer's formalism which identifies conductance with transmission \cite{landauer1957spatial, buttiker1986four, imry2002introduction}. Our aim here is to extend this formalism to  localized particle losses by modelling them by an imaginary potential.

\subsection{Effective Hamiltonian with losses}

For a system evolving according to a Hamiltonian $\hat{H}_0$ which is weakly coupled to a fast evolving environment (Born-Markov approximation), the evolution of the density matrix $\rho$ is determined by the Lindblad master equation \cite{breuer2002theory}:
\begin{equation}
\partial_t \hat{\rho} = -\frac{i}{\hbar} [\hat{H}, \hat{\rho}] - \sum_j \left(\frac 1 2 \hat{L}_j^{\dagger} \hat{L}_j \hat{\rho} + \frac 1 2 \hat{\rho} \hat{L}_j^{\dagger} \hat{L}_j -\hat{L}_j \hat{\rho} \hat{L}_j^{\dagger} \right) \label{master_equation}
\end{equation}
where $\hat{L}_j$ represent the jump operators describing the loss processes which is proportional to the square root of the scattering rate (\ref{eq:scattering_rate}).
The first two terms of the sum can be included in the commutator by defining an effective, complex-valued Hamiltonian 
\begin{equation}
\hat{H}_{\rm eff} = \hat{H}_0 - i\frac{\hbar}{2}\sum_j \hat{L}_j^{\dagger} \hat{L}_j
\end{equation}
which, for losses spatially varying along $y$, becomes:
\begin{equation}
\hat{H}_{\rm eff} = \hat{H}_0 - i\frac{\hbar\Gamma(y)}{2} \hat{\Psi}^{\dagger}(y) \hat{\Psi}(y) \label{effective_hamiltonian}
\end{equation}
where $\hat{\Psi}(y)$ is the particle annihilation operator at position $y$.

The last term in the sum of (\ref{master_equation}) represents fluctuations and is necessary to preserve fermionic commutation relations.  
Using an imaginary potential is a minimal way to capture non-Hermitian dynamics \cite{moiseyev_2011} that neglects this last term. 
This ignores the stochastic nature of the loss processes and applies well to particles described by a macroscopic wavefunction such as photons or condensed bosons \cite{carusotto_quantum_2013,barontini_controlling_2013}. 
Although unpaired fermions are not described by a macroscopic wavefunction, the last term  of (\ref{master_equation}) can still be neglected in our case since it will not contribute to macroscopic response functions measured over times very large compared to the fluctuation timescales. 
Therefore, the non-Hermitian Hamiltonian approach  (\ref{effective_hamiltonian}) should be sufficient to model the conductance measurement in a transport experiment, performed for durations much longer than the typical scattering time $ \Gamma^{-1}$.

\subsection{Transmission and loss coefficients \label{sub:tran_loss_coeff}}

In the ballistic regime, the conductance between two reservoirs at chemical potentials $\mu_L$ and $\mu_R$ only depends on the overall transmission at a given energy $E$:
\begin{equation}
 \mathcal{T}(E)= \sum_{{\rm modes}\ \mathbf{n} }\mathcal{T}_\mathbf{n} (E)
\end{equation}
where $\mathcal{T}_\mathbf{n} (E)$ is the transmission of transverse mode $\mathbf{n}$, i.e. the probability that a particle injected at one end with energy $E$ will be transmitted to the other end.

The different transverse modes of the QPC are characterized by the quantum numbers $\mathbf{n} = (n_x, n_z)$ describing the transverse wavefunction along the harmonically confined directions. We will restrict the analysis to the first transverse mode $\mathbf{n} = (0, 0)$, which is valid at low chemical potentials in the reservoirs.

The transmisssion $\mathcal{T}(E)$ is computed from the longitudinal, complex-valued, spin-dependent potential $ V_{\sigma}(y)$ that the particles experience as they travel along the QPC region. 
To this end, forward and backward scattering amplitudes associated with this complex potential are computed by solving the time-independent 1D Schr\"{o}dinger equation with Numerov's method \cite{hairer1987solving}.
Their square moduli are equal to the energy-dependent transmission $\mathcal{T}(E)$ and reflection $\mathcal{R}(E)$ respectively, and do not sum up to one since the total probability amplitude is not conserved by the non-unitary evolution.
The loss probability $\mathcal{L}(E)$ is defined as the missing probability,
\begin{equation}
\mathcal{T}(E) + \mathcal{R}(E) + \mathcal{L}(E) = 1.
\label{eq:loss_probability}
\end{equation}

\subsection{Landauer-B\"{u}ttiker formula\label{sub:LB}}

In the Landauer-B\"{u}ttiker picture presented in Fig.~\ref{fig:setup}b, transport arises from the sum of a right-moving and a left-moving current of particles with the Fermi-Dirac distributions $f_L(E)$ and $f_R(E)$ of the left and right reservoirs respectively, defined as:
\begin{equation}
f(E, \mu, T) = \frac{1}{1 + e^{\frac{E-\mu}{k_B T}}}
\end{equation}
where $\mu = \mu_{L/R}$ and $T = T_{L/R}$ are the chemical potentials and temperatures of each reservoir.  

The number of right and left movers that are transmitted through the mesoscopic channel with an energy-dependent probability $\mathcal{T}(E)$ per unit time is 
\begin{equation}
\dot{N}^{\rm trans}_{\rightarrow / \leftarrow }= \frac{1}{h} \int_{-\infty}^{+\infty} f_{L/R}(E) \mathcal{T}(E) dE.
 \label{eq:left_right_movers}\end{equation}
The total number of atoms crossing the channel per unit time is therefore:
\begin{align}
\dot{N}^{\rm trans} &= \dot{N}^{\rm trans}_{\rightarrow } + \dot{N}^{\rm trans}_{\leftarrow } \label{eq:total_trans} \\
 &=  \dot{N}^{\rm trans}_{\rightarrow } - \dot{N}^{\rm trans}_{\leftarrow } \label{eq:trans_fermi} \\
 &+ 2\dot{N}^{\rm trans}_{\leftarrow} . \label{eq:trans_below_fermi}
\end{align}
The net current is the difference between the right and left movers as in equation (\ref{eq:trans_fermi}), corresponding to the orange shaded area in  Fig.~\ref{fig:setup}b:
\begin{equation}
\dot{N}^{\rm trans}_{\rm c } = I_N = \frac{1}{h} \int_{-\infty}^{+\infty} [f_L(E) - f_R(E)] \mathcal{T}(E) dE. \label{eq:net_current}
\end{equation}
where the subscript 'c' stands for 'contributing' to transport. The rest of atoms are transmitted through the channel without contributing to the net current with a rate 
\begin{equation}
\dot{N}^{\rm trans}_{\rm nc } =2\dot{N}^{\rm trans}_{\leftarrow} \label{eq:N_nc}
\end{equation}
(with a subscript 'nc' for 'non-contributing'). This corresponds to the brown shaded area in  Fig.~\ref{fig:setup}b.

Particle transport is also associated with energy transport, leading to weak thermoelectric effects that are neglected in the rest of the paper \cite{grenier2016thermoelectric}. Assuming that the reservoirs have the same temperature $T$, conductance is obtained using Ohm's law
\begin{equation}
I_N = G \Delta \mu . \label{eq:particle_current_sm}
\end{equation}
To compute conductance in Sec.~\ref{sec:experiment}, we furthermore assume small biases $\Delta \mu$ relative to temperature $k_B T$ that simplify the expression to
\begin{equation}
G = \frac 1 {h}\int_{-\infty}^{+\infty}\mathcal{T}(E)\left(-\frac{\partial f(E,\bar{\mu}, T)}{\partial E}\right) dE \label{eq:landauer}
\end{equation}
with $\bar{\mu} = (\mu_L + \mu_R)/2$.

\subsection{Time evolution of the reservoir properties\label{sub:time_simulation}}

This Landauer-B\"{u}ttiker picture can be refined to estimate the particle and energy losses occurring during  transport.

In contrast to the net current (\ref{eq:net_current}), the absolute particle number loss is equal to the sum of the currents lost from the mesoscopic channel by photon scattering with probability $\mathcal{L}(E)$,
\begin{align}
-\frac{dN}{dt} &= \frac{1}{h} \int_{-\infty}^{+\infty} [f_L(E) + f_R(E)] \mathcal{L}(E) dE \label{eq:loss} \\
 &= \frac{1}{h} \int_{-\infty}^{+\infty} [f_L(E) - f_R(E)] \mathcal{L}(E) dE \label{eq:loss_fermi} \\
 &+ \frac{2}{h} \int_{-\infty}^{+\infty} f_R(E) \mathcal{L}(E) dE. \label{eq:loss_below_fermi}
\end{align}
Similar to equations (\ref{eq:left_right_movers}) to (\ref{eq:N_nc}), the total rate of particle losses (\ref{eq:loss} can be split into a contributing term $\dot{N}_{\rm c}^{\rm loss}$ corresponding to the atoms at the Fermi level in equation \ref{eq:loss_fermi} (orange area in Fig.~\ref{fig:setup}b) and a non-contributing term $\dot{N}_{\rm nc}^{\rm loss}$ corresponding to the atoms blelow the lowest Fermi level in equation \ref{eq:loss_below_fermi} (brown area in Fig.~\ref{fig:setup}b). 

The time evolution of each reservoir, characterized by its particle number $N_{L,R}(t)$ and internal energy $U_{L,R}(t)$, can be evaluated iteratively by (i) converting the extensive quantities $(N_{L,R}, U_{L,R})$ into chemical potential and temperature $(\mu_{L,R}, T_{L,R})$ using the equation of state of the non-interacting Fermi gas in a 3D harmonic trap \cite{su_low-temperature_2003}; (ii) inserting these quantities into the reservoir Fermi-Dirac distributions $f_{L,R}$ to compute the time-derivatives of particle number and energy in each reservoir:
\begin{align}
\frac{dN_{L,R}}{dt} &= \frac{1}{h} \int_{-\infty}^{+\infty} f_{R,L}(E) \mathcal{T}(E) dE \nonumber \\
&- \frac{1}{h} \int_{-\infty}^{+\infty} f_{L,R}(E) [\mathcal{T}(E) + \mathcal{L}(E)] dE \\
\frac{dU_{L,R}}{dt} &= \frac{1}{h} \int_{-\infty}^{+\infty} E \cdot f_{R,L}(E) \mathcal{T}(E) dE \nonumber \\
&- \frac{1}{h} \int_{-\infty}^{+\infty} E \cdot f_{L,R}(E) [\mathcal{T}(E) + \mathcal{L}(E)] dE;
\end{align}
and (iii) updating the particle number and energy after a numerical time step using Euler's method.

\section{Experimental procedure \label{sec:loss_mechanisms}}

In this section, we present the characteristics of the experiment and of the conductance measurements. We also justify why the scattering of a photon, normally associated with a large momentum kick, can be modeled by a local particle loss without increasing the total energy of the system. 

\subsection{Experimental parameters \label{sub:experimental_geometry}}

We prepare atoms in the geometry depicted in Fig~\ref{fig:setup}a of an atomic QPC consisting of two reservoirs connected by a 1D region where the atomic density is tunable.

We start by producing a degenerate cloud of $^6$Li atoms in a balanced mixture of $\ket\downarrow$ and $\ket\uparrow$ (the first and third lowest hyperfine states), with a typical temperature of $T=\SI{66(12)}{\nano\kelvin}$ and $N=1.1(1)\cdot 10^5$ atoms per spin state. Before starting the transport experiment, the magnetic field is ramped to a value of $\SI{568}{\gauss}$ and $\SI{574}{\gauss}$ for Fig~\ref{fig:losses} and Fig~\ref{fig:losses_and_conductance}-\ref{fig:dsgm} where the scattering length between the two components is $a=0(7)\,a_0$ and $a=91(7)\,a_0$, respectively. The atoms are thus very weakly interacting and atom-atom scattering is expected to be weak, even in the 1D region of the QPC. 

The cloud is then shaped into two reservoirs by projecting the different optical potentials, see Fig~\ref{fig:setup}a. The vertical (resp. horizontal) confinement is provided by a beam propagating along $x$ (resp. $z$) with a waist along $y$ of $w_z=\SI{30.2}{\micro\meter}$ (resp. $w_x=\SI{5.9}{\micro\meter}$) with maximum confinement frequency $\nu_z=\SI{9.03(5)}{\kilo\hertz}$ (resp. $\nu_y=\SI{14.0(6)}{\kilo\hertz}$). 
The length of the QPC is mainly defined by the shortest waist of the constriction beams, namely $w_x$. 
The mean chemical potential of the reservoirs is typically $\mu_\text{res} = (\mu_L + \mu_R)/2 = k_B \cdot \SI{0.23}{\micro\kelvin}$.  The density in and around the 1D region is tuned using an attractive gate beam of waist $w_g=\SI{31.8(3)}{\micro\meter}$ and of maximum potential $V_g$.  
This increases the local value of the chemical potential relevant for understanding transport to $V_g + \mu_\text{res}$.
Inside the QPC, we add a near-resonant light beam whose frequency can be tuned between the two resonances for states $\ket{\uparrow}$ and $\ket{\downarrow}$ of the D$_2$ line for $\sigma^-$ polarized light (see Fig~\ref{fig:theory_detuning_high_B}ab). This beam is shaped into a Gaussian profile with a waist of $w_s =\SI{2.0(1)}{\micro\meter}$ thanks to a digital micromirror device that allows for aberration correction and precise positioning inside the constriction \cite{zupancic2016ultra}. Its power of $P_s = \SI{20(6)}{\pico\watt}$ corresponds to a peak intensity of $I_s = 2 P_s/\pi w_s^2 = \SI{3(1)}{\watt/\meter^2} = 0.13(4)\,I_\text{sat}$.  
For equal and opposite detunings from the two resonances $\bar{\delta} = 0$, this corresponds to a dipole potential of $V_s = V_\uparrow = -V_\downarrow = k_B \cdot\SI{330(98)}{\nano\kelvin} $ and a photon absorption rate $\Gamma = 3.1(9)\cdot 10^3\,\SI{}{\second^{-1}}$.

Neglecting the spatial variations of the attractive gate beam, the potential landscape $ V_{\sigma}(y)$ defined in subsection \ref{sub:tran_loss_coeff} consists of: 
\begin{enumerate}
\item a space-dependent zero-point energy due to the $x$- and $z$-confinement
\begin{equation}
V_0(y) =\frac 1 2  h\nu_z e^{-y^2/w_z^2}+\frac 1 2  h\nu_x e^{-y^2/w_x^2} \label{eq:zpe}
\end{equation}
where $\nu_{x,z}$ are the maximum confinement frequencies and $w_{x,z}$ the waists along $y$ of the beams providing this confinement;
\item a spin-dependent potential $V_\sigma(y)$ defined by (\ref{eq:dipole_potential}) and proportional to the near-resonant beam intensity $I(y) = I_s e^{-2y^2/w_s^2}$:
\begin{equation}
V_\sigma(y) = \epsilon_\sigma V_s e^{-2y^2/w_s^2} \label{eq:v_sigma}
\end{equation}
with $\epsilon_\uparrow = +1$ and $\epsilon_\downarrow = -1$;
\item an imaginary potential $iV_{\rm loss}(y) = -i\hbar \frac{\Gamma(y)}{2}$ introduced in the effective Hamiltonian (\ref{effective_hamiltonian}), describing losses due to the scattering rate $\Gamma(y)$ defined by (\ref{eq:scattering_rate}) that is proportional to $I(y)$.
\end{enumerate}

At the end of an experimental cycle, the atomic density of the cloud is recorded using absorption imaging. The density profile is then fitted to the equation of state of the non-interacting Fermi gas to extract the atom number, temperature and chemical potential of each reservoir. 

\subsection{Transport measurement}

Transport experiments are performed by introducing an initial atom number between the two reservoirs and letting the system relax to equilibrium via the flow of particles of each internal state through the ballistic channel. 
Typically, we prepare for each spin state atom number differences of $\Delta N(0) = 45(3) \cdot 10^3$. 
As demonstrated in previous works, in the absence of dissipation, this system is very well described using Landauer-B\"uttiker theory \cite{krinner_observation_2015}. 

When the temperature difference between the reservoir is zero, \emph{i.e.} $\Delta T = 0$, the particle current through the QPC is linear in the chemical potential between the reservoirs following equation~(\ref{eq:particle_current_sm}).
The linear approximation is valid for the weak interaction strengths considered here \cite{krinner_mapping_2016}. They also ensure that spin drag is negligible and that biases, currents, and transport coefficients can be treated independently for each spin $\sigma \in \{\uparrow, \downarrow\}$ (omitted in the rest of the subsection).

The chemical potential for each reservoir $r \in \{L, R\}$ and each spin can be furthermore expanded to first order around the atom number at $t = 0$, $d\mu_r = dN_r / \kappa_r$, where $\kappa_r$ is the compressibility of one reservoir.
Equation (\ref{eq:particle_current_sm}) can then be simplified to a closed first order differential equation in the atom number difference $\Delta N$.
This difference is an exponentially decreasing function of time, showing that the atomic QPC is the analogue of an RC circuit for neutral atoms, with a time constant
\begin{equation}
\tau = G \left(\frac{1}{\kappa_{\rm L}}+\frac{1}{\kappa_{\rm R}} \right) \approx \frac {2G}{\kappa}\ ,
\label{eq:decay_time_RC}
\end{equation}
where $\kappa$ is the reservoir compressibility at global equilibrium for a mean atom number $\bar{N} = (N_L + N_R)/2$ and temperature $T$ computed from the trap geometry and the non-interacting Fermi gas properties. 
The timescale $\tau$ is extracted from the atom number difference $\Delta N / N = (N_L - N_R)/(N_L + N_R)$ at $t=\SI{0}{\second}$ and after a fixed transport time $t=\SI{4}{\second}$, which in turn yields the conductance $G$.

Using thermodynamical variables such as $T$, $\mu$ or $\kappa$ is possible as long as the reservoirs can be described by thermal states, which is valid when the scattering time in the reservoirs $\tau_s$ is small compared to the characteristic timescale of the transport $\tau$.
The scattering time can be approximated by $\tau_s = 1/\bar{n} \sigma v_F$, where $\bar{n}$ is the peak density and $v_F$ the Fermi velocity taken at the trap center, and $\sigma$ is the interatomic scattering cross-section. 
For a scattering length of $a=91 \,a_0$, $\tau_s = \SI{0.08}{\second} $ while $\tau$ is on the order of several seconds for a conductance around $1/h$. In the absence of atom-atom scattering for  $a=0(7) \,a_0$, even though this assumption should formally break down, we expect the reservoirs to remain approximated by thermal states since atom number variations due to transport and losses are smaller than 10\% per reservoir. Such an agreement between observables in thermalized and non-thermalized regimes was already noted in our previous experimental work \cite{krinner_observation_2015}, and is furthermore supported by theoretical studies on the validity of the Landauer formalism both in presence of incoherent baths or in a complete microcanonical picture \cite{chien_landauer_2014}.

\subsection{Loss mechanisms}

When an atom scatters a near-resonant photon, it gets a large kinetic energy corresponding to the two momentum kicks associated with the absorption and spontaneous reemission of the photon. It is therefore not lost immediately but travels through the cloud of trapped atoms and can interact with them. In this subsection, we show that an atom that underwent a scattering event can be considered as lost for moderate interaction strengths between the two component of the gas. This requires to study (i) the single particle dynamics after a photon scattering event as well as the subsequent scattering with (ii) atoms inside the wire and (iii) atoms in the reservoir.

Atoms that absorb a photon re-emit it spontaneously at a rate $\Gamma_0 = \SI{36.9}{\micro\second^{-1}}$ during which they move by less than one nanometer.
The imparted recoil energy $E_R = (h/\lambda)^2/2m = k_B \cdot \SI{3.54}{\micro\kelvin}$  
is smaller than the potential barrier imposed by the beams defining the QPC, equal to $k_B \cdot \SI{48}{\micro\kelvin}$ in the $z$-direction and $k_B \cdot \SI{7}{\micro\kelvin}$ in the $x$-direction.
Atoms therefore do not escape the QPC laterally and are projected to a superposition of transverse QPC modes, described by the Lamb-Dicke parameter $\eta_z = \sqrt{E_R/h \nu_z} = 2.86(1)$ for a recoil momentum transfer along $z$ and $\eta_x = \sqrt{E_R/h \nu_x} = 2.29(4)$ along $x$.
They are however directed towards the reservoirs along the non-confined direction $y$.

Once the atom has scattered a photon, it is projected into an excited state of the 2D harmonic oscillator whose quantum number is $n_{x,z} \simeq \eta_{x,z}^2$ on average and can still scatter with another low energy atom in the wire.
With strong harmonic confinement in two directions of space, the s-wave scattering properties can be modified by confinement induced resonances \cite{olshanii1998atomic}. Using \cite{moore2004scattering}, we can compute the transmission probability associated to the scattering event of an atom in $(n_x, n_z)$ with an atom in the ground state of the two transverse harmonic oscillator at a relative momentum of $ k_R = 2\pi/\lambda$. The probability that the atoms are transmitted without changing their oscillator and momentum states is equal to $T = 1-6.1\cdot 10^{-4} $, which has to be exponentiated by the number of potential atom-atom scattering events in the channel. There is therefore less than $0.4\%$ chance that such a collision happens with any atom in the wire, hence scattering events between the energetic particle and the atoms in the channel can be neglected.

When an atom at the recoil velocity enters the three-dimensional reservoirs, the relevant quantity to consider is its mean free path $\ell=1/\bar{n} \sigma $. For the largest interaction strength considered here, its value is $\ell = \SI{3.26(5)}{\milli\meter}$, much larger than the reservoir size of approximately $\SI{0.2}{\milli\meter}$.
Since the recoil energy is larger than the depth $V_{\rm trap} = k_B \cdot \SI{0.55}{\micro\kelvin}$ of the optical trap defining the reservoirs, scattered atoms eventually escape the system and do not contribute to a global energy increase for the weak interactions. 

Therefore, for weak s-wave interactions, an atom which has scattered a photon can be considered as lost and the formalism of section \ref{sec:lossy_landauer} applies.

\begin{figure}
    \centering
    \includegraphics[scale=1]{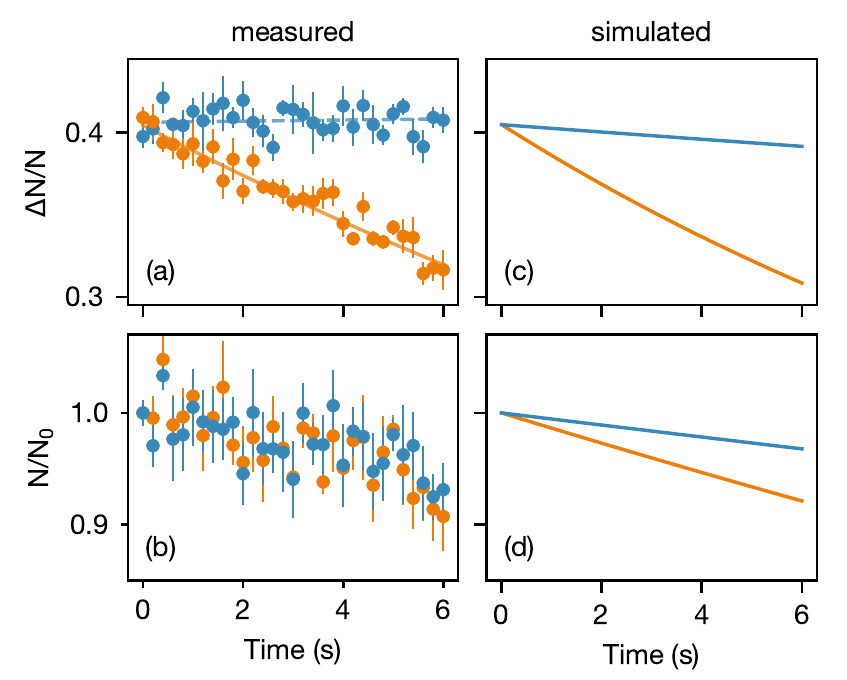}
    \caption{\textbf{Measured and simulated evolution of the relative atom number imbalance and of the atom number.}
    (\textbf{a}) Experimental relative imbalance and 
    (\textbf{b}) normalized atom number for spin-$\ket{\downarrow}$ (orange) and spin-$\ket{\uparrow}$ (blue), obtained for a mean chemical potential $V_g + \mu_\text{res} = k_B \cdot \SI{0.61(2)}{\micro\kelvin}$,  with a near-resonant beam power $P_s = \SI{20(6)}{\pico\watt}$ corresponding to an intensity $I_s = 0.13(4)\,I_\text{sat}$, where $I_\text{sat}$ is the $D_2$-line saturation intensity and at a scattering length $a = 0(7)\,a_0$.
    The initial atom numbers for each spin state are $N_{0,\downarrow} = 117 \cdot 10^3$ and  $N_{0,\uparrow} = 110 \cdot 10^3$. Here and in the following, error bars correspond to the standard error of the mean of 3 measurements. 
    (\textbf{c}) Simulation results for the relative atom number difference $\Delta N/N$ and (\textbf{d}) the normalized atom number with spin-dependent potential $V_s = k_B \cdot \SI{0.25}{\micro\kelvin}$, gate potential $V_g = k_B \cdot \SI{0.38}{\micro\kelvin}$, temperature $T = \SI{60}{\nano\kelvin}$ and initial conditions $N_{0,\downarrow / \uparrow} = 115 \cdot 10^3$ and $\Delta N_0/N_0 = 0.41$.
    Losses occur mostly for $\ket{\downarrow}$-atoms. 
    }
    \label{fig:losses}
\end{figure}

\section{Experimental investigation of transport with losses\label{sec:experiment}}

We now compare the measurements to the models of Sec.~\ref{sec:polarizability} and \ref{sec:lossy_landauer}. First, we compare the experimental time evolution of the total atom number and relative imbalance for a fixed value of the detuning to the simulation. From this evolution, we can extract the conductance of the QPC and demonstrate that the expected conductance plateaus remain visible even at large local chemical potential, with a value renormalized by the loss probability. We then vary the detuning of the near-resonant tweezer to show the validity of the Landauer-B\"uttiker model even when the tweezer is brought on resonance with one of the two spin states. Last, varying the position of the tweezer in the channel generates losses proportional to the local atomic density, allowing to map it in and around the QPC.

\subsection{Atom number evolution in the reservoirs}

In a first measurement, we study the evolution of the atom number in each reservoir as a function of time both experimentally and numerically.

With the experimental setup and parameters described in subsection \ref{sub:experimental_geometry}, the photon scattering rates are equal 
for the two internal states while the potentials are of equal and opposite magnitude, as illustrated in Fig.~\ref{fig:theory_detuning_high_B}ab. The relative imbalance $\Delta N/N = (N_L-N_R)/(N_L+N_R)$ and the normalized atom number $N/N_0= (N_L+N_R)/(N_L(t=0)+N_R(t=0))$ are recorded over $6\,$s at a magnetic field of $568\,$G where the s-wave scattering length is $a = 0(7)\,a_0$. The results, presented in Figs.~\ref{fig:losses}ab, illustrate that the spin-dependent potential acts as a repulsive barrier for the $\ket{\uparrow}$ state: the fitted current $I_\uparrow = -19 \pm 85 \,\SI{}{\second^{-1}}$ vanishes while the one for the $\ket{\downarrow}$ has a finite value of  $I_\downarrow = 833 \pm 98\,\SI{}{\second^{-1}}$ corresponding to a conductance of $G_{\downarrow}=0.45(4)/h$. In spite of a maximal photon scattering rate $\Gamma_s = 2.3(8)\cdot 10^3\,\SI{}{\second^{-1}}$ at the center of the tweezer, the atom losses are moderate since they represent less than 10\% of the total atom number. 

This experiment is reproduced by a numerical simulation following section \ref{sub:time_simulation} using the parameters obtained experimentally (near-resonant beam intensity, initial atom number, initial imbalance, temperature). 
The numerical results shown in Figs.~\ref{fig:losses}cd are largely consistent with the experimental results, indicating that for moderate values of the conductance, neglecting fluctuations, extending the Landauer-B\"uttiker model and integrating the time evolution of the cloud properties as in section~\ref{sec:lossy_landauer} are a valid approach. The simulation also indicates that losses should be smaller for $\ket{\uparrow}$-atoms which are blocked by the spin-dependent potential and therefore have lower densities in the dissipative region. This effect nevertheless remains elusive in the experimental data, partly because of the uncertainty in the measured atom numbers.

\begin{figure}
    \includegraphics[scale=1]{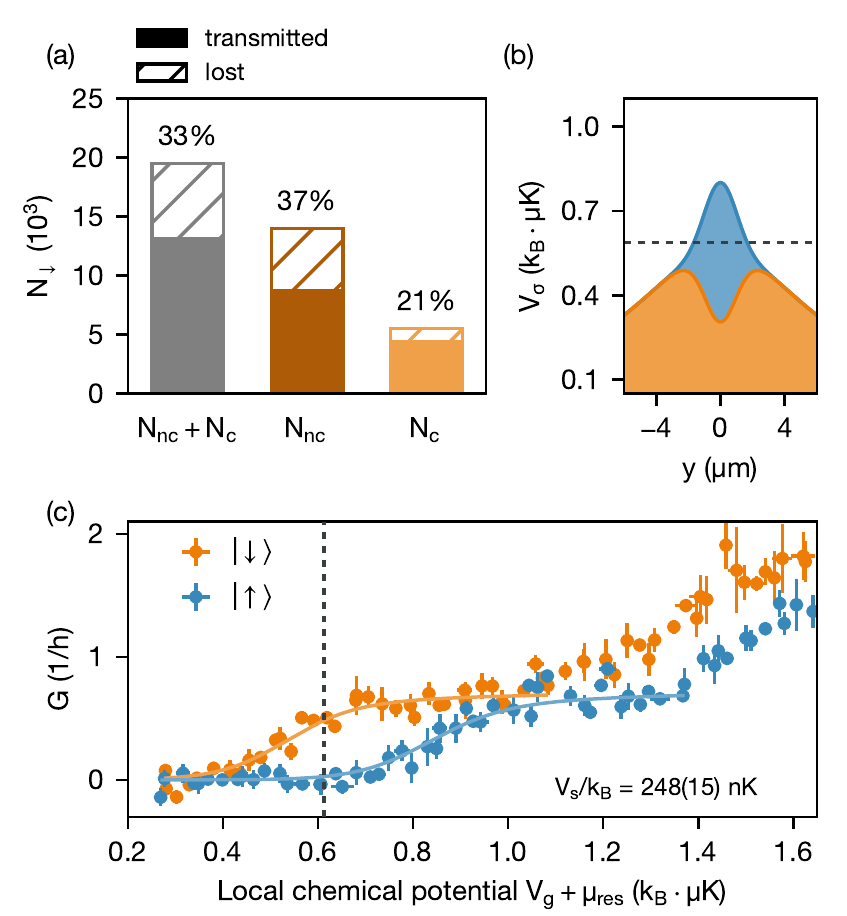}
    \caption{\textbf{Conductance plateaus are preserved at a lossy QPC. }
    (\textbf{a}) Breakdown of the simulated numbers of lost and transmitted atoms as a function of their energy using the parameters of Fig.~\ref{fig:losses} integrated over a transport time of $\SI{4}{\second}$. This highlights one reason for the robustness of transport observables. The total number of lost (resp. transmitted) atoms [first bar of the chart] can be decomposed into two parts, shown in filled (resp. hatched) region: (i) those which do not contribute to transport [equation (\ref{eq:trans_below_fermi}) and (\ref{eq:loss_below_fermi}), brown region in Fig.~\ref{fig:setup} and second bar of the chart] and (ii) those which could contribute to transport [equation (\ref{eq:trans_fermi}) and (\ref{eq:loss_fermi}), orange region in Fig.~\ref{fig:setup} and third bar of the chart], see text for details. The relative fraction of lost atoms in each category, indicated on top of each bar, shows that the losses are less important for the atoms participating to transport (due to their higher velocities) than for the other atoms, hence conductance is weakly affected by the losses.
    (\textbf{b}) Quasi-1D potentials along the transport direction $y$ for each spin state $V_{\sigma}(y)$.
    The chemical potential of Fig.~\ref{fig:losses} and subfigure (a) is indicated as dashed lines.
    (\textbf{c}) Conductance $G$ of each spin state at scattering length $a = 91(7)\,a_0$ versus local chemical potential $V_g+\mu_{\textrm{res}}$ with a near-resonant beam intensity of  $I_s = 0.13(4)\,I_\text{sat}$.
    Fits by a Landauer model are shown as solid curves and indicate a spin-dependent potential of $V_s = k_B \cdot \SI{0.25(2)}{\micro\kelvin}$.
    }
    \label{fig:losses_and_conductance}
\end{figure}

\subsection{Preserving the conductance plateaus at a lossy QPC}

In Section \ref{sec:lossy_landauer}, we have showed that transport observables are sensitive only to scattering at energies close to the Fermi level which concerns a small fraction of all atoms subject to near-resonant light. Conductance is therefore expected to be robust against losses. 
As illustrated in Fig.~\ref{fig:setup}b (brown areas), most of the losses actually concern atoms that are below the Fermi surface (\ref{eq:loss_below_fermi}) and therefore do not affect conductance. 

This is verified by integrating the simulation results of the Fig.~\ref{fig:losses} over $t = \SI{4}{\second}$ of transport time.  
We thus obtain the number of atoms 
transmitted or lost and participating or not to transport (subscript 'c' and 'nc') 
\begin{equation}
N_{\rm c/nc}^{\rm trans/lost}=\int_0^{t}\dot{N}_{\rm c/nc}^{\rm trans/lost} dt'
\end{equation}
  using the quantities defined in subsection~\ref{sub:LB} and \ref{sub:time_simulation}. 
We then extract the total number of particle participating to transport $N_{\rm c} = N_{\rm c}^{\rm lost} + N_{\rm c}^{\rm trans} $ and non-participating to transport $N_{\rm nc} = N_{\rm nc}^{\rm lost} + N_{\rm nc}^{\rm trans} $.
The value of these quantities and their sum is represented in Fig~\ref{fig:losses_and_conductance}b for the $\ket{\downarrow}$ state: while 33\% of the particles flowing through the dissipative region are lost, only 21\% of the particles contributing to net transport are dissipated due to their larger velocities than the non-contributing particles. 

Recording conductance as a function of local chemical potential demonstrates that plateaus are still visible at the lossy QPC, as shown in Fig.~\ref{fig:losses_and_conductance}b. We fit the conductances of both states with the Landauer model (solid curves in Fig.~\ref{fig:losses_and_conductance}b) described in section \ref{sec:lossy_landauer}.
This yields an experimental value of the spin-dependent potential of $V_s=k_B\cdot \SI{0.25(2)}{\micro\kelvin}$, compatible with the theoretical value of  $k_B\cdot \SI{0.29(11)}{\micro\kelvin}$ for an intensity of $\SI{3(1)}{\watt/\meter^2}$

In a Landauer picture valid for weak interactions, these losses contribute to decreasing the conductance by the scattering probability. 
This probability is computed to be 21\% in Fig.~\ref{fig:losses_and_conductance}a and is consistent with the decrease of the conductance plateau from $G = 0.84(1)/h$ to $G = 0.72(3)/h$.

\begin{figure}
    \includegraphics[scale=1]{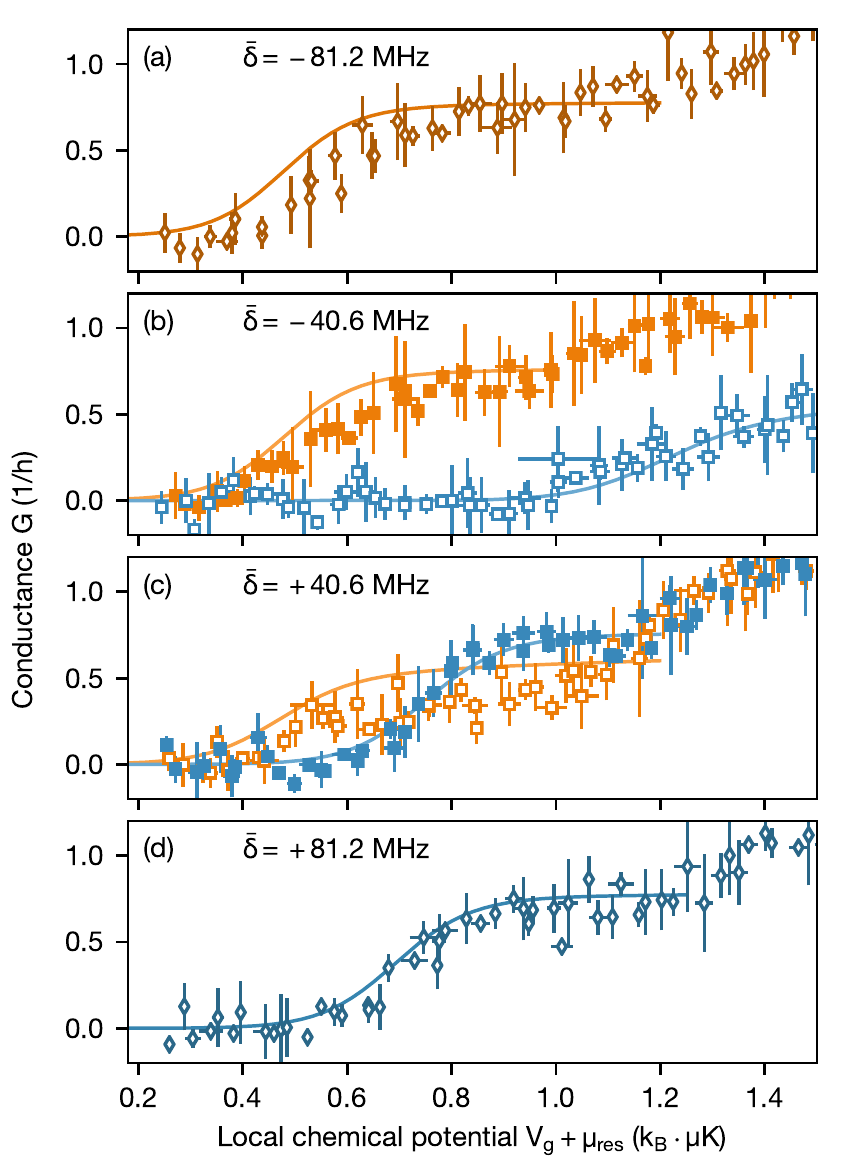}
    \caption{\textbf{Validity of the Landauer model for large dissipation.}
     Conductance $G$ at a scattering length $a = 91(7)\,a_0$  for spin-$\ket{\downarrow}$ (orange) and spin-$\ket{\uparrow}$ (blue) with  
     (\textbf{a}) near-resonant light on resonance with $\ket{\uparrow}$ at detuning  $\bar{\delta} = -\SI{81.3}{\mega\hertz}$ for $\ket{\downarrow}$;
    (\textbf{b}) at detuning $\bar{\delta} = \SI{-40.6}{\mega\hertz}$ for both $\ket{\downarrow}$ (orange) and $\ket{\uparrow}$ (blue);
    (\textbf{c}) at detuning $\bar{\delta} = \SI{40.6}{\mega\hertz}$;
    (\textbf{d}) tweezer light on resonance with $\ket{\downarrow}$ at detuning  $\bar{\delta} = \SI{81.3}{\mega\hertz}$ for $\ket{\uparrow}$. 
    Solid curves show a Landauer prediction using the fit parameters of Fig.~\ref{fig:losses_and_conductance}c extended to different detunings. The different detunings at which the conductance is measured are indicated by dashed grey lines in Fig.~\ref{fig:theory_detuning_high_B}ab.
    }
    \label{fig:detuning}
\end{figure}

\subsection{Varying the loss rates}

To explore the validity range of this Landauer model with losses presented in section \ref{sec:lossy_landauer}, we extend our conductance measurements to different tweezer detunings $\bar{\delta}$ relative to the mean resonance frequency $(\nu_\uparrow + \nu_\downarrow)/2$ at fixed intensity $I_s = 0.13(4)\,I_\text{sat}$ and at an interaction strength of $ 91(7)\,a_0$.

As shown in Figs.~\ref{fig:theory_detuning_high_B}ab, this affects both spin-dependent dipole potential $V_\sigma$ and photon scattering rate $\Gamma_\sigma$, and allows to change the latter by more than three orders of magnitude. We explore detunings ranging from the tweezer being resonant with the $\ket{\uparrow}$ at $\bar{\delta}=\SI{-81.2}{\mega\hertz}$ in Fig.~\ref{fig:detuning}a to the  $\ket{\downarrow}$ resonance at $\bar{\delta}=\SI{81.2}{\mega\hertz}$ in Fig.~\ref{fig:detuning}d.
Bringing the tweezer on resonance with one of the two states leads to its entire loss after $\SI{4}{\second}$, while the other non-resonant state still displays quantized conductance (Fig.~\ref{fig:detuning}a and d).
The applicability of the Landauer theory highlights that the coupling between spins is negligible at this scattering length.

In addition, tuning the frequency by $\bar{\delta} = -\SI{40.6}{\mega\hertz}$ towards the resonance of $\ket{\uparrow}$ leads to a shift of the conductance curve towards higher chemical potentials (Fig.~\ref{fig:detuning}b, blue), since the repulsive potential barrier and scattering rate for that state are increased.
Meanwhile, the conductance of $\ket{\downarrow}$ (Fig.~\ref{fig:detuning}b, orange) approaches the one measured in the absence of near-resonant light.

The reverse trend is observed with a detuning $\bar{\delta} = +\SI{40.6}{\mega\hertz}$, where the conductance of $\ket{\downarrow}$ is clearly reduced due to increased losses (Fig.~\ref{fig:detuning}c).
These measurements show good agreement with the previous Landauer model without having to add any fit parameters.

The good agreement demonstrates the applicability of the Landauer-B\"uttiker formula over a wide range of dissipation strengths. 

\begin{figure}
    \includegraphics{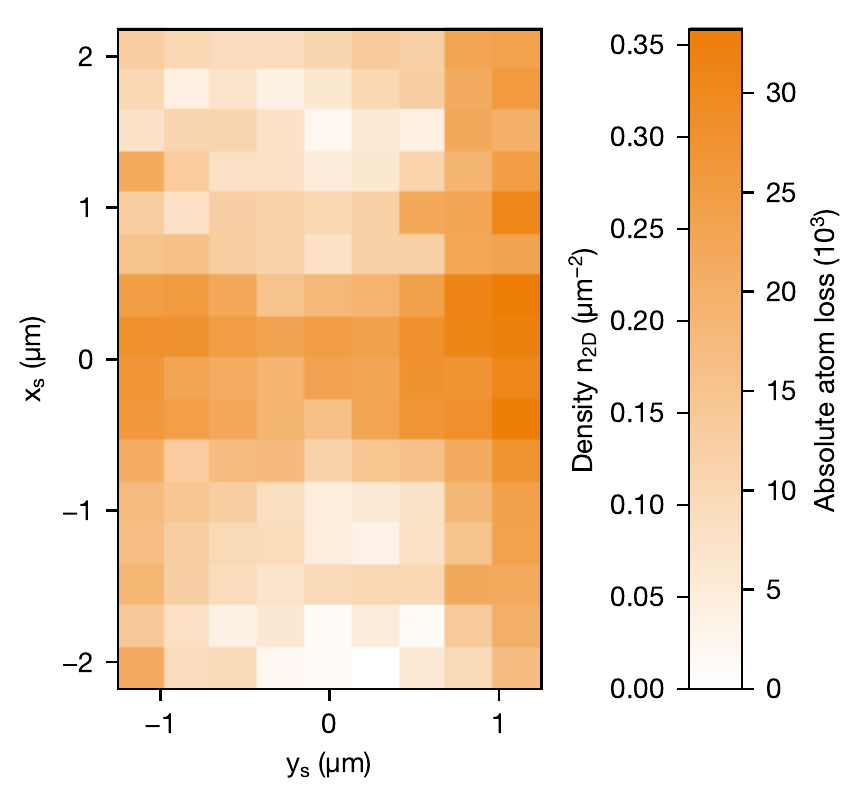}
    \caption{\textbf{Dissipative scanning gate microscopy.}
    Integrated atom loss in state $\ket{\downarrow}$ after a time $\Delta t = \SI{0.63}{\second}$ as a function of the near-resonant beam position $(x_s, y_s)$ and corresponding two-dimensional density $n_\text{2D}$, showing a high-resolution map of the QPC.
    The near-resonant beam has a detuning of $\bar{\delta} = \SI{40}{\mega\hertz}$, a narrower waist of $w_s = \SI{1.02(5)}{\micro\meter}$ and a power $P_s = \SI{4(1)e1}{\pico\watt}$, corresponding to a peak intensity of $I_s = \SI{24(8)}{\watt/\meter^2}$ and therefore a large photon absorption rate $\Gamma = \SI{9(3)e4}{\second^{-1}}$ for state $\ket{\downarrow}$.
    }
    \label{fig:dsgm}
\end{figure}

\subsection{Scanning gate microscopy with losses}

We have so far studied the effect of dissipation in the transport structure, developing an extension to the Landauer model; now we can further use the dissipative beam as a density probe. The near-resonant tweezer is corrected for aberrations using a digital micromirror device in Fourier configuration. A discretized grating therefore controls the phase front of the beam which determines its position inside the QPC at the sub-micron level. We take advantage of this precise positioning by recording the total atom loss as a function of the tweezer position to infer the local atomic density. We perform this experiment with a vanishing atom imbalance between the reservoirs. 

To model the situation, we assume that the atomic density is two-dimensional and time-independent. These assumptions hold provided that the density variations along the $z$ direction are small compared to the Rayleigh length of the near-resonant beam $z_R = \pi w_s^2/\lambda = \SI{4.9(3)}{\micro\meter}$, and that the atom losses remain small relative to the total atom number. 

The atom losses integrated over the time interval $\Delta t$ during which photon scattering occur can be written as:
\begin{align}
N(0) - N(t) &= \Delta t \int dx dy \Gamma(x, y) n_\text{2D}(x, y) dx dy \\
\text{with } \Gamma(x, y) &= \Gamma_s e^{-2 [(x-x_s)^2 + (y-y_s)^2]/w_s^2}
\end{align}

Losses are proportional to the atomic density convolved with the Gaussian intensity profile of the near-resonant beam centered at position $(x_s, y_s)$. In the limit where its Gaussian waist $w_s$ is small with respect to the variations of the atomic density, it can be approximated by a Dirac function and the local density is given by:
\begin{equation}
n_\text{2D}(x_s, y_s) = \frac{2}{\pi w_s^2} \frac{N(0) - N(t)}{\Gamma_s \Delta t}.
\end{equation}

Repeating measurements of atom losses for different positions of the near-resonant beam allows us to retrieve a direct map of the two-dimensional density, as shown in Fig.~\ref{fig:dsgm}.
This measurement is conceptually similar to scanning probe microscopy in solid-state physics, and complements previous imaging techniques for quantum gases using a focused electron beam \cite{gericke_high-resolution_2008} or a conservative optical potential in a transport geometry \cite{hausler_scanning_2017}.

\subsection{Conclusion}

Adding a near-resonant beam at an atomic QPC leads to different potentials and scattering rates for the different spin states. These value can be computed from the atomic polarizability and used to express the transmission and loss probabilities of each particle traveling through the point contact. This required extending the Landauer-B\"uttiker formalism to a situation with losses which reproduces the measurement of the conductance, where a plateau remains visible although its value is decreased with respect to the quantum of conductance $1/h$. It is also possible to integrate the results of this model to express the time evolution of the atom number and relative imbalance between the reservoirs of the two-terminal geometry. Finally, the atom losses can be related to the atomic density at the position of the near-resonant tweezer which can therefore act as ``dissipative scanning gate microscope'' for ultracold gases.

The ability to engineer dissipation in a transport experiment opens the possibility to study the competition between losses and coherent transport by investigating the continuous Zeno effect \cite{froml2019fluctuation} or the modification of transport through a mesoscopic, dissipative lattice. Ultracold atoms also allow to vary the $s$-wave interaction strength up to point where a paired superfluid is formed. There, characteristic signatures of transport through a tunnel barrier can also be strongly influenced by the presence of dissipation \cite{damanet2019controlling}.

\section*{Acknowledgments}
We thank T.~Giamarchi, L.~Glazman, H.~Moritz, H.~Ott and A.-M.~Visuri for helpful discussions; and J.-P.~Brantut, R.~Citro, M.~Landini, and K.~Viebahn for their critical reading of the manuscript.
We acknowledge the Swiss National Science Foundation (Project n$^{\circ}$ 182650 and NCCR-QSIT) and ERC advanced grant TransQ (Project n$^{\circ}$ 742579) for funding.
L.C. is supported by ETH Zurich Postdoctoral Fellowship, Marie Curie Actions for People COFUND program and EU Horizon 2020 Marie Curie TopSpiD (Project n$^{\circ}$ 746150).

\bibliographystyle{apsrev4-1}
% \bibliography{paper}

%merlin.mbs apsrev4-1.bst 2010-07-25 4.21a (PWD, AO, DPC) hacked
%Control: key (0)
%Control: author (72) initials jnrlst
%Control: editor formatted (1) identically to author
%Control: production of article title (-1) disabled
%Control: page (0) single
%Control: year (1) truncated
%Control: production of eprint (0) enabled
%

\end{document}